\documentclass[aps,prd,reprint,showpacs,floatfix,a4paper]{revtex4-1}

\usepackage[english]{babel}
\usepackage{graphicx}
\usepackage{array}
\usepackage{dcolumn}
\usepackage{bm}
\usepackage{amssymb}
\usepackage{amsmath}
\usepackage{amsfonts}
\usepackage{amsbsy}
\usepackage{hyperref}
\usepackage{breakurl}

\begin{document}

\title{Probability distribution for the relative velocity of colliding particles in a relativistic 
classical gas}

\author{Mirco Cannoni}

\affiliation{Departamento de F\'isica Aplicada, Facultad de Ciencias
Experimentales, Universidad de Huelva, 21071 Huelva, Spain}

\begin{abstract}

We find the probability density function $\mathcal{P}(V_{\texttt{r}})$ of the 
relativistic relative velocity for two colliding particles in a 
non-degenerate relativistic gas. 
The distribution reduces to Maxwell distribution for the relative 
velocity in the non-relativistic limit. 
We find an exact formula for the mean value $\langle V_{\texttt{r}}\rangle$. 
The mean velocity tends to the Maxwell's value in the 
non-relativistic limit and to the velocity of light in the ultra-relativistic limit.
At a given temperature $T$, when at least for one of the two particles the ratio of the rest energy 
over the thermal energy $m c^2/k_B T$ is smaller than 40 
the Maxwell distribution is inadequate.

\end{abstract}

\pacs{03.30.+p,11.80.-m,05.20.-y}
	
\date{October 16, 2013 }

\maketitle

\section{Introduction}

In many fields of physics and astrophysics one has to study reaction rates in a system that 
can be considered, to a good approximation, a classical non-relativistic gas in equilibrium. 
In the gas can be present different species of particles.
We consider two species with masses $m_1$ and $m_2$
and number densities $n_{1,2}$, the number of particles per unit volume.
For a given process with total cross section $\sigma$, 
the number of reactions per unit time per unit volume,
the reaction rate, is given by $R=n_1 n_2 \sigma v_{{r}}$, where 
\begin{equation}
v_{{r}}=\rvert \boldsymbol{v}_1 - \boldsymbol{v}_2 \rvert 
\label{vRelNR}
\end{equation}
is the relative velocity between two particles with velocities $\boldsymbol{v}_1$ and $\boldsymbol{v}_2$.
The cross section is in general function of the relative velocity but we will not write explicitly
the dependence. 
As it is well known, at a given temperature $T$~\cite{units}, the absolute velocity of the particles follows 
the Maxwell distribution $f_M (v)=(2/\pi)^{1/2}(m/T)^{3/2}v^2 \exp(-mv^2/(2T))$.
The thermally averaged reaction rate then is
\begin{equation}
\langle R \rangle =n_1 n_2 \int d\boldsymbol{v}_1 d\boldsymbol{v}_2 f_M ({v}_1) 
f_M({v}_2) \sigma 
{v}_{{r}}.
\label{NR_rate}
\end{equation}
By changing variables 
from the velocities $\boldsymbol{v}_1$, $\boldsymbol{v}_2$ to the velocity of 
the center of mass $\boldsymbol{v}_c $ and the relative velocity $\boldsymbol{v}_r$, 
one finds the standard expression for the thermal averaged rate,
\begin{equation}
\langle R\rangle =n_1 n_2 \int_{0}^{\infty} dv_{{r}}  F_{\text{M}}(v_{{r}})\sigma v_{{r}} ,
\label{NR_rate_final}
\end{equation}
where
\begin{equation}
F_{{M}}(v_{{r}})=\sqrt{\frac{2}{\pi}}\left( \frac{\mu}{T}\right)^{3/2} 
{v}^2_{{r}}\, e^{-\mu\frac{v_{{r}}^2}{2T}}
\label{Maxwell_vrel}
\end{equation}
is the distribution of the relative velocity. Equation (\ref{Maxwell_vrel}) has the
same form of the Maxwell distribution for the absolute velocity but
with the reduced mass $\mu =m_1 m_2/(m_1 +m_2)$ in place of $m$
and $v_{{r}}$ in place of $v$. 

If the colliding particles are relativistic 
corrections to Eq.~(\ref{NR_rate})  can be important.
On the other hand, conceptually, both  the relative velocity (\ref{vRelNR}) and the Maxwell 
distribution (\ref{Maxwell_vrel}) are not compatible with the fact that $v_r$ for two massive particles 
must be smaller than velocity of light $c$ in every inertial frame, while 
the relative velocity between two massless particles 
and between a massless and 
a massive particle is always equal to the velocity of light.

It is thus interesting to ask if a probability distribution for the relative velocity 
compatible with the principles of special relativity exists.
In this paper we show that such a distribution exists
and that $F_{{M}}(v_{{r}})$ is just its non-relativistic limit.

Let us remind first how the previous discussion of the non-relativistic reaction rate 
is reformulated in a Lorentz invariant way.
The relativistic relative velocity is~\cite{Landau2,Cercignani} 
\begin{flalign}
V_\texttt{r}=
\frac{\sqrt{(\boldsymbol{v}_1 - \boldsymbol{v}_2)^2 - (\boldsymbol{v}_1 
\times 
\boldsymbol{v}_2)^2}}
{1-\boldsymbol{v}_1 \cdot \boldsymbol{v}_2}.
\label{vRelRel}
\end{flalign} 
This expression is symmetric in the two velocities in any frame and have all the required properties.
In the non-relativistic limit $V_\texttt{r}$ reduces to (\ref{vRelNR}).
The Lorentz invariant rate is~\cite{Landau2,Cercignani}
\begin{equation}
\mathcal{R}= n_1 n_2 \frac{p_1\cdot p_2}{E_1 E_2} \sigma V_\texttt{r},
\label{Rate_rel}
\end{equation}
where $p_i=(E_i ,\boldsymbol{p}_i)$, $E_i= \sqrt{\boldsymbol{p}^2_i +m^2_i}$, $i=1,2$, are the four-momentum
of the colliding particles.

A relativistic non-degenerate gas in equilibrium is described 
by the relativistic generalization of the Maxwell distribution, the 
J\"{u}ttner distribution~\cite{juttner,DeGroot,csernai,Cercignani}.
The normalized  momentum  distribution is given by
\begin{equation}
f_{J}(\boldsymbol{p})= \frac{1}{4\pi m^2 T K_2 (x)} e^{-\frac{u \cdot p}{T}}.
\label{F_J}
\end{equation}
Here and in what follows, $K_n (x)$ are  modified Bessel functions of the second kind of order $n$,
and $u$ a time-like four-velocity of the gas such that $u\cdot u=1$. 
Averaging the rate (\ref{Rate_rel}) with the J\"{u}ttner distribution (\ref{F_J}), the relativistic analogous 
of Eq.~(\ref{NR_rate}) hence is 
\begin{equation}
\langle \mathcal{R}\rangle=n_1 n_2 \int \frac{d^3 \boldsymbol{p}_1}{E_1} \frac{d^3 \boldsymbol{p}_2}{E_2} 
 f_{J}(\boldsymbol{p}_1) f_{J}(\boldsymbol{p}_2)\,
(p_1\cdot p_2)  \sigma   V_\texttt{r}.
\label{Rel_averaged_rate}
\end{equation}
This is our starting point.

\section{Probability distribution for the relative velocity}

In Eq.~(\ref{Rel_averaged_rate}) the integrand is manifestly Lorentz invariant. In order to simplify the calculation, 
we can choose the so-called Lorentz local rest frame~\cite{DeGroot,Cercignani,csernai} where the four velocity 
of the gas is $u=(1,\boldsymbol{0})$. Hence, \textit{(1)} we show that
\begin{flalign}
\int \frac{d^3 \boldsymbol{p}_1}{E_1}  \frac{d^3 \boldsymbol{p}_2}{E_2}
{p_1\cdot p_2} f_{J}(\boldsymbol{p}_1) f_{J}(\boldsymbol{p}_2) 
\equiv \int^{1}_{0} dV_{\texttt{r}} \mathcal{P}_{\texttt{r}}(V_{\texttt{r}}) =1;
\nonumber
\end{flalign}
\textit{(2)} we give the explicit expression for $\mathcal{P}_{\texttt{r}}(V_{\texttt{r}})$;
\textit{(3)} we verify $\mathcal{P}_{\texttt{r}}(V_{\texttt{r}})$ in the non-relativistic limit
reduces to Eq.~(\ref{Maxwell_vrel}).

\textit{(1)}
Introducing the ratios $x_{i}=m_{i}/T$ we have
\begin{equation}
\Phi=\frac{\int \frac{d^3 \boldsymbol{p}_1}{E_1} \frac{d^3 \boldsymbol{p}_2}{E_2} 
p_1\cdot p_2 e^{-\frac{E_1 + E_2}{T}}}
{(4\pi T)^2 \prod_{i} m^2_i K_2 (x_i)}=\frac{\mathcal{N}}{\mathcal{D}}.
\label{Phi}
\end{equation}
The integrand in the numerator depends on $\theta$, the angle between $\boldsymbol{p}_1$ and $\boldsymbol{p}_2$, trough
the scalar product $p_1 \cdot p_2$.
Passing in polar coordinates in momentum space, $d^3 \boldsymbol{p}_1=4\pi |\boldsymbol{p}_1|^2 
d|\boldsymbol{p}_1|$,
$d^3 \boldsymbol{p}_2 = 2\pi |\boldsymbol{p}_2|^2 d|\boldsymbol{p}_1|d\cos\theta$, 
the integration over the angle gives
\[
\int^{1}_{-1} d\cos\theta (E_1 E_2-|\boldsymbol{p}_1||\boldsymbol{p}_2|\cos\theta )=2 E_1 E_2.
\]
The numerator is thus 
$\mathcal{N}=\prod_i \int d^3 \boldsymbol{p}_i e^{-\frac{E_i}{T}}=\mathcal{D}$
%
because of the normalization of the J\"{u}ttner distribution~(\ref{F_J}). It follows that $\Phi=1$.  

\textit{(2)} To find the explicit expression for $\mathcal{P}_{\texttt{r}}(V_{\texttt{r}})$
it is convenient to follow Refs.~\cite{GG} and change variables from $E_1$, $E_2$, $\cos\theta$ to $Y=E_1 +E_2$, $Z=E_1- E_2$ and the Mandelstam invariant $s=(p_1 +p_2)^2$.
Defining  
\begin{equation}
p'= \frac{\sqrt{s-(m_1 + m_2)^2} \sqrt{s-(m_1 - m_2)^2} }    {2\sqrt{s}}.
\end{equation}
and $M=m_1 +m_2$,  with $p_1 \cdot p_2 =[s-(m^2_1 +m^2_2)]/2$, we obtain 
\begin{equation}
\Phi=\frac{\int_{M^2}^{\infty} ds [s-(m^2_1 +m^2_2)] p' K_1 (\sqrt{s}/T)}
{4 T \prod_{i} m^2_i K_2 (x_i) } .
\label{Phi_s}
\end{equation}
Introducing the Lorentz factor associated with $V_{\texttt{r}}$,
\begin{equation}
\gamma_{\texttt{r}}=\frac{1}{\sqrt{1-v^2_{\texttt{r}}}},
\end{equation}
and the relative velocity (\ref{vRelRel}) in terms of  $p_{1,2}$,
\[
V_{\texttt{r}}=\frac{\sqrt{(p_1 \cdot p_2)^2 -m^2_1 m^2_2}}{p_1 \cdot p_2},
\]
we express $s$ as a function of $\gamma_{\texttt{r}}$, 
\begin{equation}
s=(m_1 - m_2)^2 +2 m_1 m_2 (1+\gamma_{\texttt{r}}),
\label{s_gamma}
\end{equation}
and change variable from $s$ to $\gamma_{\texttt{r}}$.
The integral, after some algebra, can be cast in the compact form
\begin{equation}
\Phi=\frac{X\int_{1}^{\infty} d\gamma_{_{\texttt{r}}}
\gamma_{_{\texttt{r}}} \sqrt{\frac{\gamma^2_{_{\texttt{r}}}-1}{\gamma_{_{\texttt{r}}}+\varrho}}
K_1 (\sqrt{2}X\sqrt{\gamma_{_{\texttt{r}}} +\varrho}) }{\sqrt{2} \prod_{i} K_2 (x_i) }, 
\label{Phi_gamma}
\end{equation}
where we have defined the abbreviations
\begin{equation}
X=\sqrt{x_1 x_2},\;\;\varrho=\frac{m^2_1 +m^2_2}{2m_1 m_2}=\frac{x^2_1 +x^2_2}{2x_1 x_2}.
\label{X_ro}
\end{equation}
We now change again variable from $\gamma_{_{\texttt{r}}}$ to $V_{\texttt{r}}$
with the differential 
$d\gamma_{\texttt{r}}= \gamma^2_{\texttt{r}}
\sqrt{\gamma^2_{\texttt{r}}-1}dV_{\texttt{r}}$,
\begin{equation}
\Phi=\frac{X\int_{0}^{1} dV_{\texttt{r}}
\gamma^3_{_{\texttt{r}}} \frac{\gamma^2_{_{\texttt{r}}} -1}{\sqrt{\gamma_{{\texttt{r}}} +\varrho}}
K_1 (\sqrt{2}X\sqrt{\gamma_{{\texttt{r}}} +\varrho})}
{\sqrt{2} \prod_{i} K_2 (x_i) }.
\label{Phi_v}
\end{equation}
From (\ref{Phi_v}) we finally read the expression for  $\mathcal{P}_{\texttt{r}}(V_{\texttt{r}})$:
\begin{equation}
\begin{array}{l}
\mathcal{P}_{\texttt{r}}(V_{\texttt{r}})=
\frac{X}{\sqrt{2} \prod_{i} K_2 (x_i) }
 \frac{\gamma^3_{_{\texttt{r}}} (\gamma^2_{_{\texttt{r}}} -1)}{\sqrt{\gamma_{{\texttt{r}}} +\varrho}}
K_1 (\sqrt{2}X\sqrt{\gamma_{{\texttt{r}}} +\varrho}).
\end{array}
\label{P_v}
\end{equation}
In the 'diagonal' case $m_1 =m_2 =m$,  we have $x_1=x_2$ and $X=x$, $\varrho=1$ thus  (\ref{P_v}) becomes
\begin{flalign}
\begin{array}{l}
\mathcal{P}^{d}_{\texttt{r}}(V_{\texttt{r}})=\frac{x}{\sqrt{2}K^2_2 (x)}
\gamma^3_{_{\texttt{r}}} \frac{\gamma^2_{_{\texttt{r}}} -1}{\sqrt{\gamma_{{\texttt{r}}} +1}} 
K_1 (\sqrt{2}x\sqrt{\gamma_{{\texttt{r}}} +1}).
\end{array}
\label{P_v_ii}
\end{flalign}
\begin{figure*}[htp!]
\includegraphics*[scale=0.6,bb=0 0 288 288,clip=true]{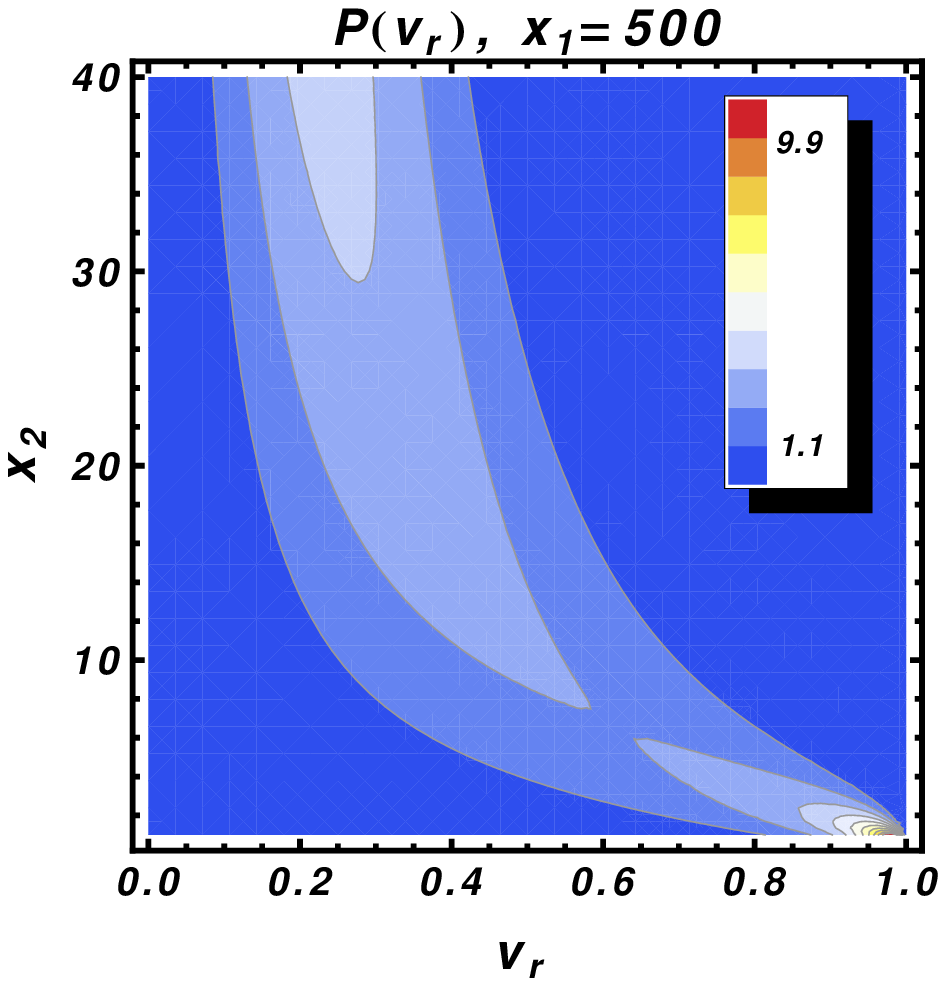}
\includegraphics*[scale=0.6,bb=0 0 288 288,clip=true]{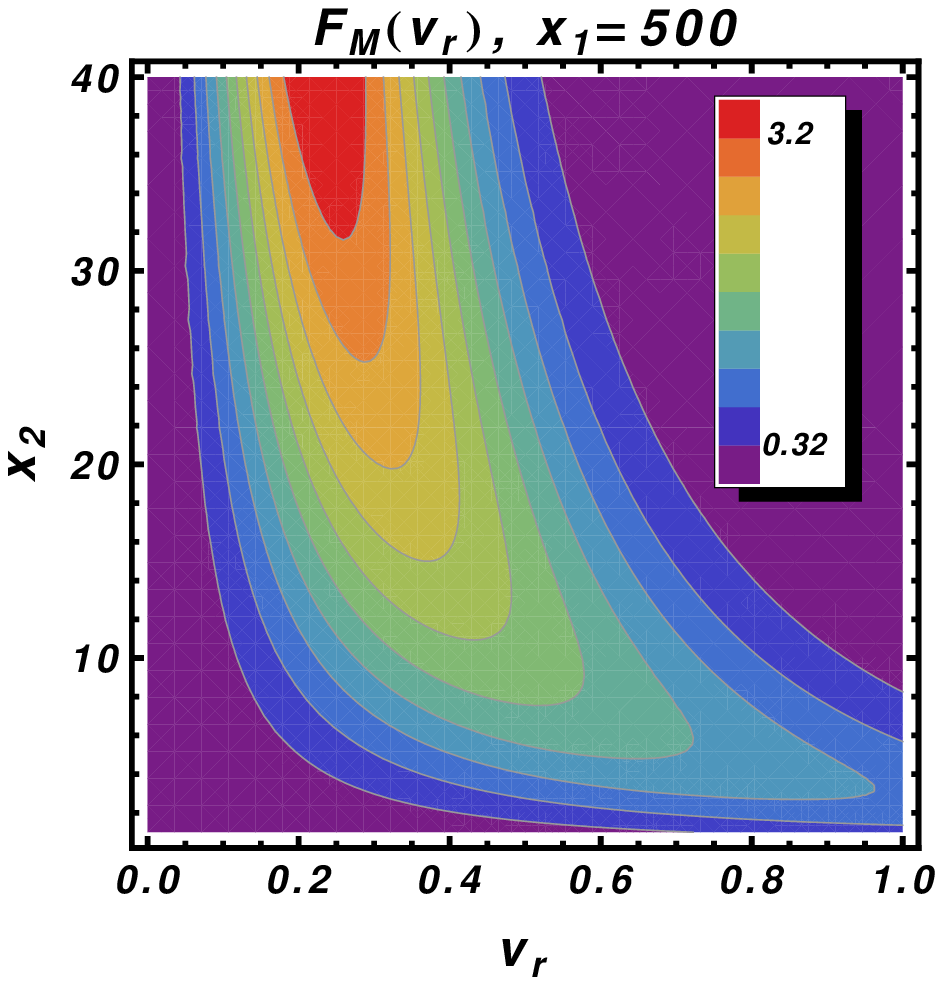}\\
\includegraphics*[scale=0.43,bb=0 0 288 288,clip=true]{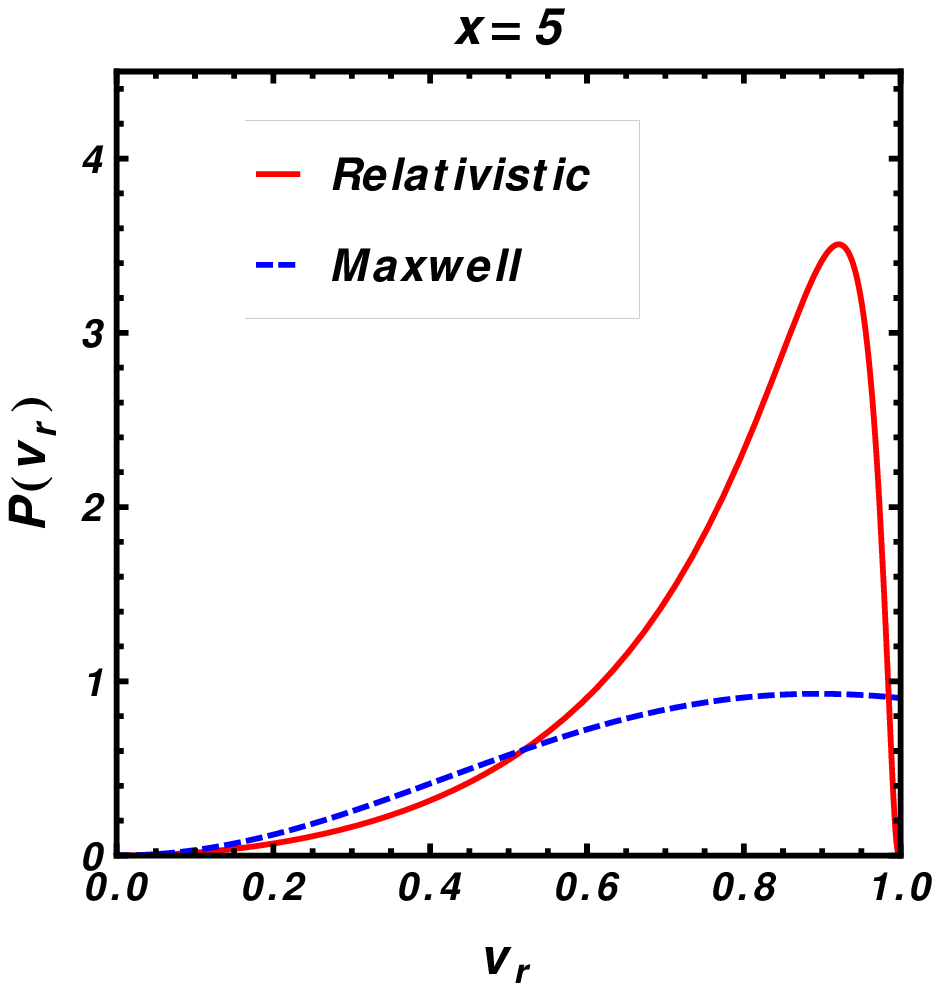}
\includegraphics*[scale=0.43,bb=0 0 288 288,clip=true]{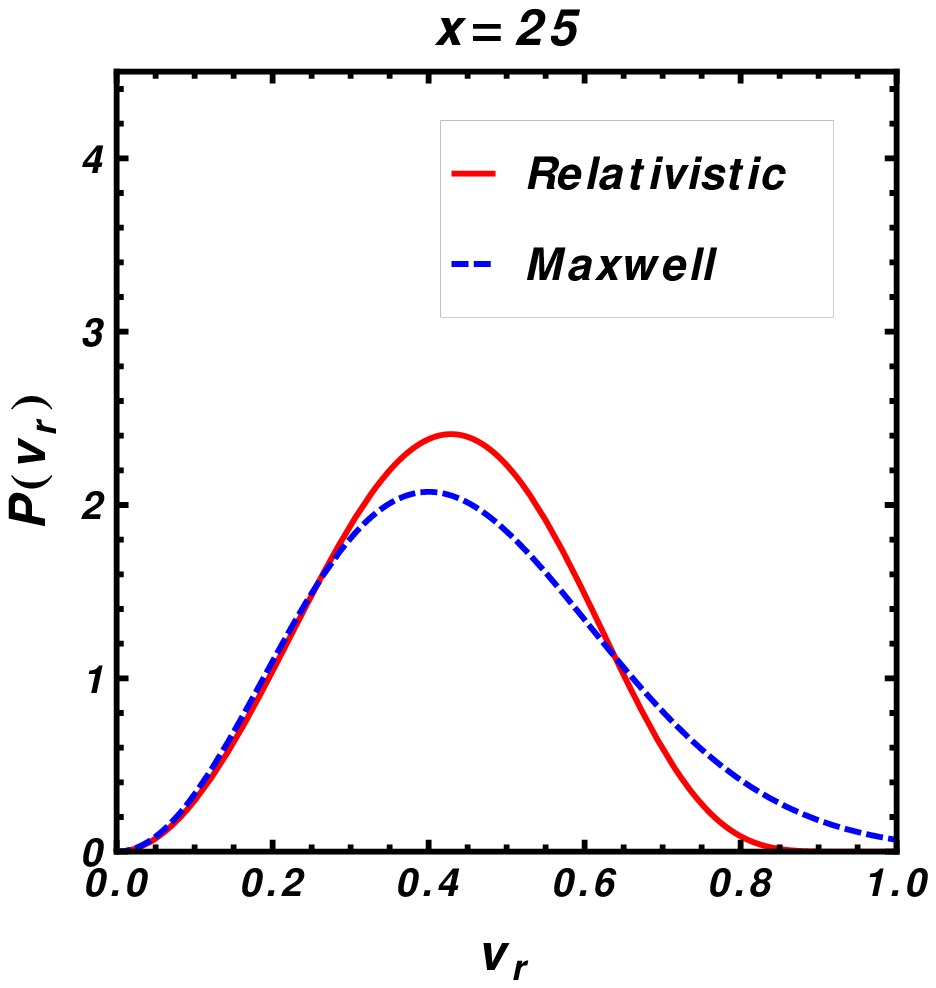}
\includegraphics*[scale=0.43,bb=0 0 288 288,clip=true]{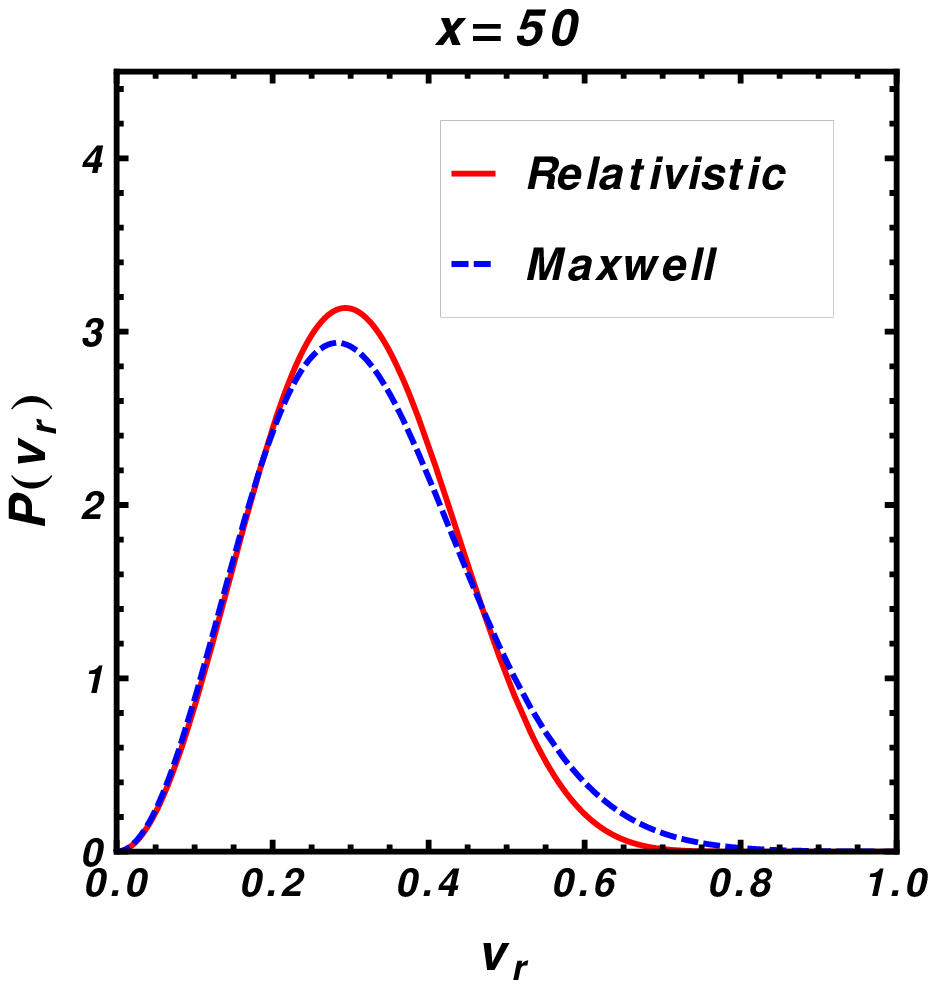}
\includegraphics*[scale=0.43,bb=0 0 288 288,clip=true]{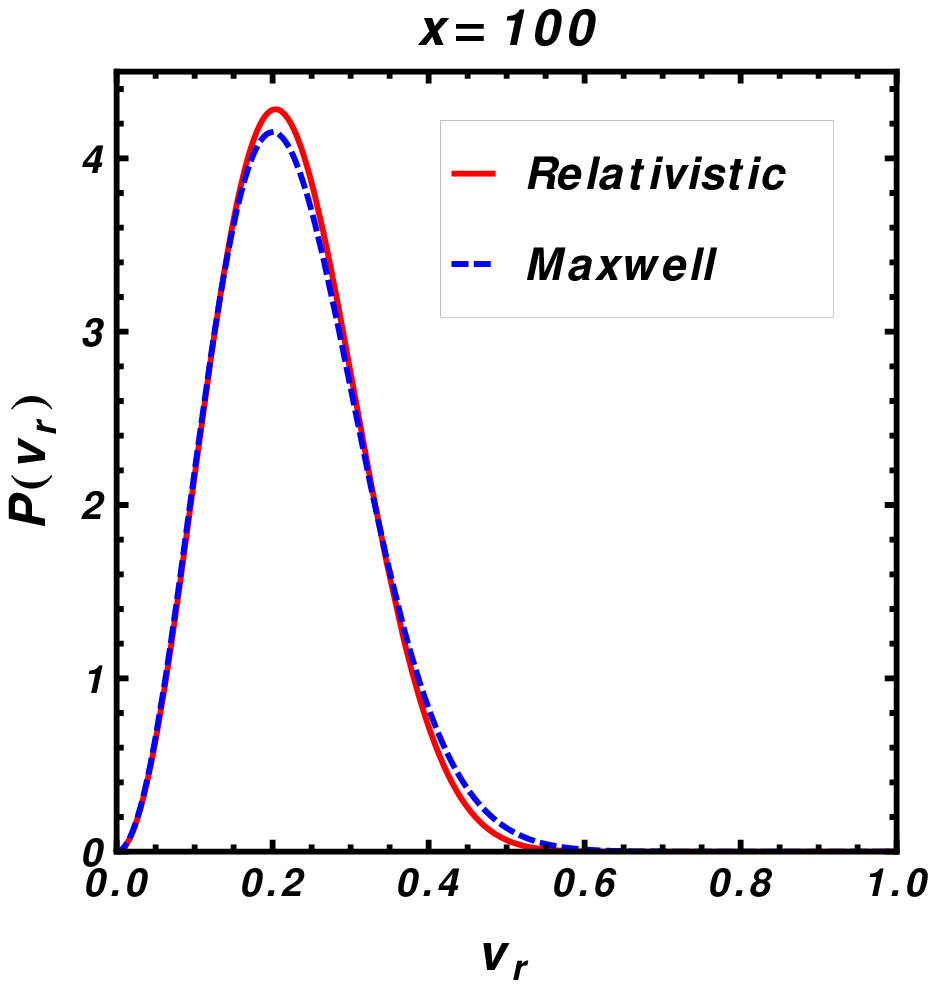}
\caption{\textit{Top-left panel}: contours of the 
relativistic distribution (\ref{P_v}) in the plane $(V_{\texttt{r}}$, $x_2$).
\textit{Top-right panel}: contours of the Maxwell distribution (\ref{Maxwell_vrel}).
The parameter $x_i=m_i/T$ for the first particle is fixed to
the non-relativistic value of  $x_1 =500$.
\textit{Bottom panels}:
The relativistic distribution, Eq.~(\ref{P_v_ii}), red lines, and the 
Maxwell distribution (\ref{Maxwell_vrel}), blue lines, as a function of the relative velocity
for increasing values of $x=m/T$. Here $m_1 =m_2 =m$.
}
\label{Fig1}
\end{figure*}

\textit{(3)}
In the non-relativistic limit $V_{\text{r}}\sim v_r\ll 1$ and 
$x_{1,2}\gg 1$.
It is useful to note the following relations between the parameters of the distribution: 
\begin{flalign}
&\sqrt{2} X \sqrt{1+\varrho}=\frac{M}{T}=x_1 +x_2 \equiv \alpha ,
\label{a}
\\
&\frac{X}{\sqrt{2}\sqrt{1+\varrho}}=\frac{\mu}{T}=\frac{x_1 x_2}{x_1+ x_2}\equiv \beta,
\label{b}
\end{flalign} 
and that for  $x\gg 1$ the asymptotic behavior  of the modified Bessel function is $K_n (x)\sim 
e^{-x}\sqrt{{\pi}/{(2x)}}$.
To the lowest order in $v^2_r$ and  $X$ we have
\begin{flalign}
&\frac{X}{\sqrt{2} \prod_{i} K_2 (x_i) }\sim \frac{\sqrt{2}}{\pi}X^2 e^\alpha,
\label{fac1}
\\
&\gamma^3_{_{\texttt{r}}} \frac{\gamma^2_{_{\texttt{r}}} -1}{\sqrt{\gamma_{{\texttt{r}}} +\varrho}}
\sim \frac{v^2_{r}}{\sqrt{1+\varrho}}=\sqrt{2}\frac{\beta}{X}{v^2_{r}},
\label{fac2}
\\
&K_1 (\sqrt{2}X\sqrt{\gamma_{{\texttt{r}}} +\varrho})\sim
\sqrt{\frac{\pi}{2}} \frac{\sqrt{\beta}}{X}
e^{-\alpha}
e^{-\beta\frac{v^2_r}{2}}.
\label{fac3}
\end{flalign}
Multiplying the Eqs.~(\ref{fac1}), (\ref{fac2}), (\ref{fac3}) we obtain the Maxwell distribution 
(\ref{Maxwell_vrel}).

%
In actual calculations it is convenient to work with different variables rather than $V_{\texttt{r}}$. 
From Eq.~(\ref{Phi_s}) we can read the distribution as a function of $s$ 
and the respective diagonal form:
\begin{flalign}
\begin{array}{l}
\mathcal{P}_{\texttt{r}}(s)=\frac{1} {4 T \prod_{i} m^2_i K_2 (x_i) }
[s-(m^2_1 +m^2_2)] p' K_1 (\frac{\sqrt{s}}{T}),
\end{array}
\label{P_s}
\\
\begin{array}{l}
\mathcal{P}^{d}_{\texttt{r}}(s)=\frac{1}{8Tm^4K^2_2 (x) }(s-2m^2)\sqrt{s-4m^2}K_1(\frac{\sqrt{s}}{T}).
\end{array}
\label{P_s_ii}
\end{flalign}
From Eq.~(\ref{Phi_gamma}) the distribution as function of $\gamma_{\texttt{r}}$ is
\begin{flalign}
&\begin{array}{l}
\mathcal{P}_{\texttt{r}}(\gamma_{_{\texttt{r}}})=
\frac{X}{\sqrt{2} \prod_{i} K_2 (x_i) }
\gamma_{_{\texttt{r}}} 
\sqrt{\frac{\gamma^2_{_{\texttt{r}}}-1}{\gamma_{_{\texttt{r}}}+\varrho}}
K_1 (\sqrt{2}X\sqrt{\gamma_{_{\texttt{r}}} +\varrho}),
\end{array}
\label{P_gamma}
\\
&\begin{array}{l}
\mathcal{P}^{d}_{\texttt{r}}(\gamma_{\texttt{r}})=\frac{x}{\sqrt{2}K^2_2(x)} \gamma_{\texttt{r}}
\sqrt{\gamma_{\texttt{r}}-1} 
K_1 (\sqrt{2}x\sqrt{\gamma_{\texttt{r}}+1}).
\end{array}
\label{P_gamma_ii}
\end{flalign}
\begin{figure*}[t!]
\includegraphics*[scale=0.58,bb=0 0 288 288,clip=true]{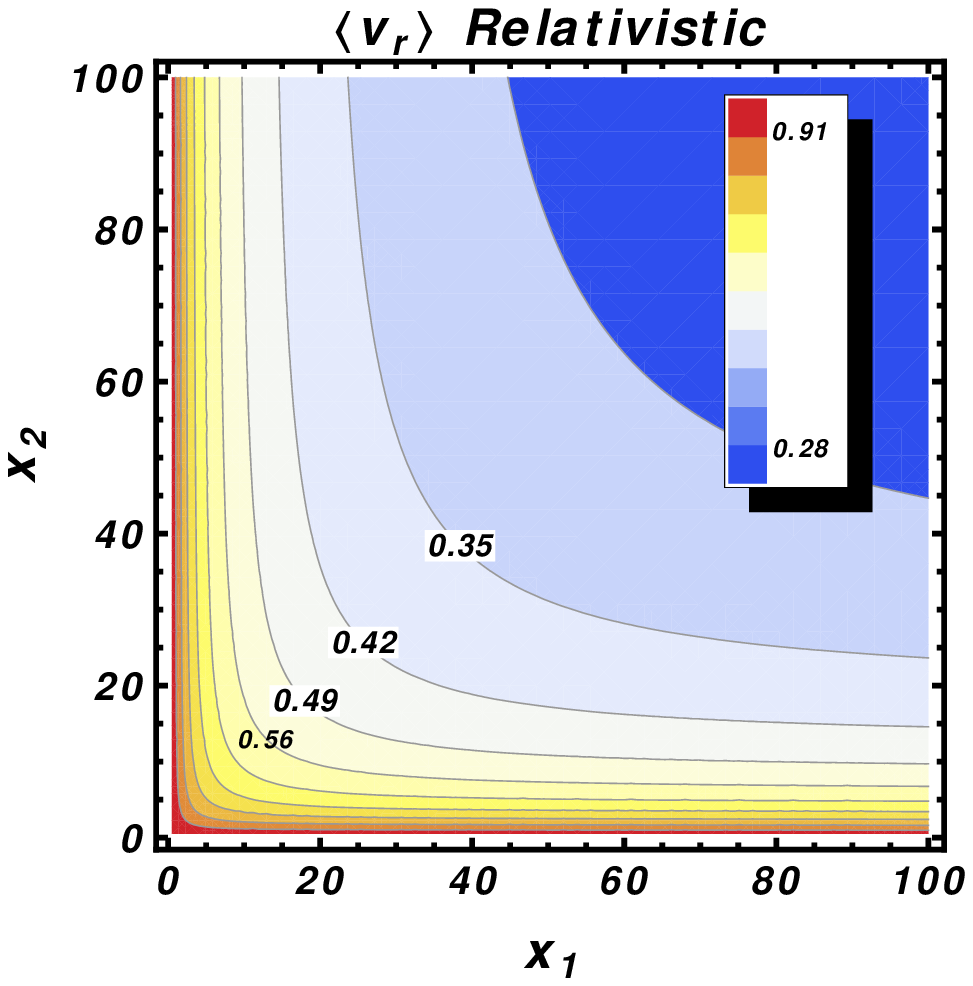}
\includegraphics*[scale=0.58,bb=0 0 288 288,clip=true]{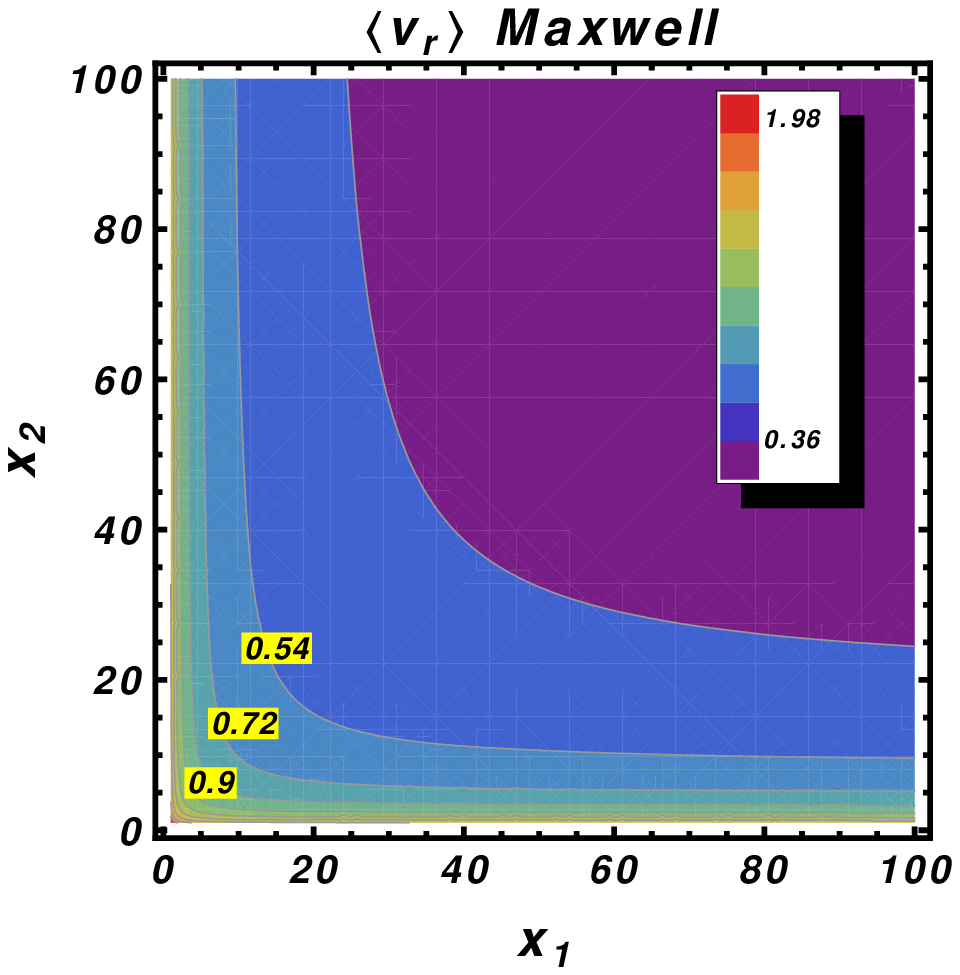}
\includegraphics*[scale=0.58,bb=0 0 288 288,clip=true]{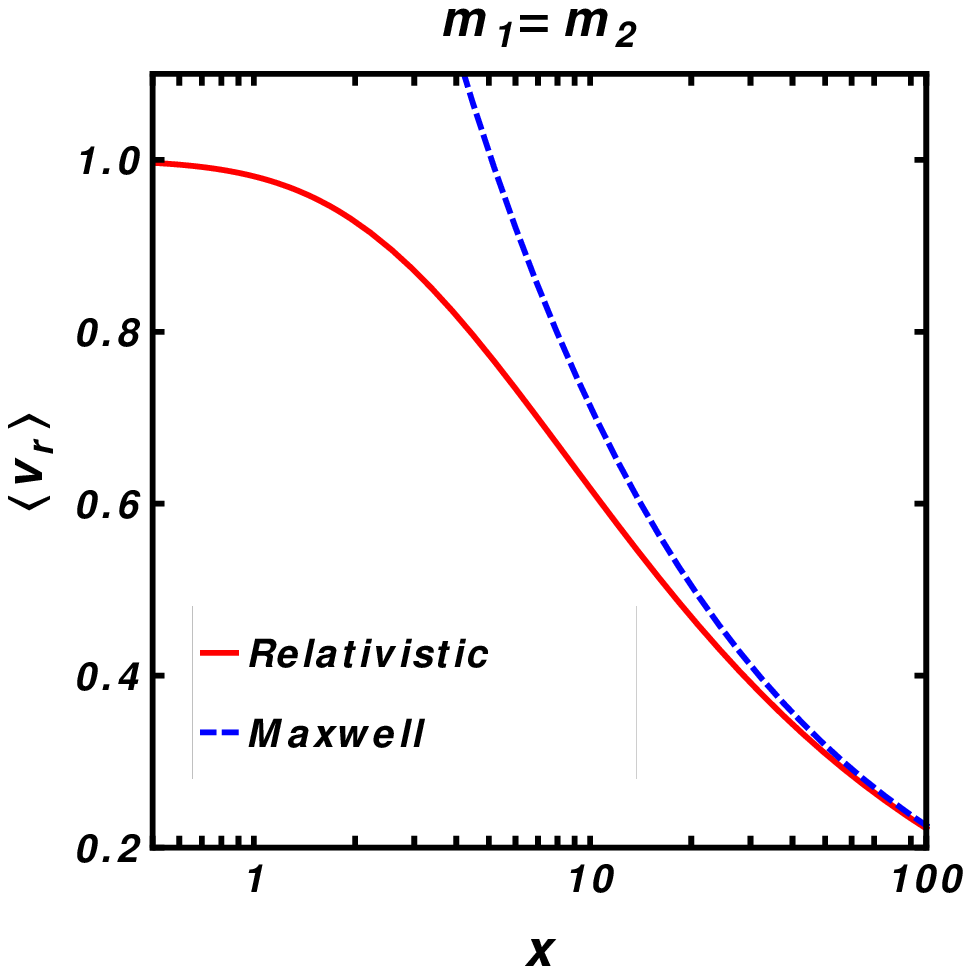}
\caption{\textit{Left panel}: contours of the mean  value of the relativistic relative velocity, Eq.~(\ref{Vmean_x1_x2})
in the plane ($x_1$, $x_2$).
\textit{Central panel}: contours in the same plane of the mean value of the relative velocity of the 
Maxwell distribution Eq.~(\ref{Vmean_Maxwell_x1_x2}).
\textit{Right panel}: Eq.~(\ref{v_rel_media}) as a function of $x=m/T$ , red line.
The blue line is the  Maxwell value (\ref{Vmean_Maxwell_x1_x2}) with
$m_1 =m_2 =m$.}
\label{Fig2}
\end{figure*}

Both $\mathcal{P}_{\texttt{r}}(V_{\texttt{r}})$ and $F_M(v_{{r}})$ depend on the masses and the 
temperature through the ratios $x_i$ and are symmetric for the exchange $x_1\leftrightarrow x_2$.
For a fixed temperature of the gas, when the masses are such that both $x_1$ and $x_2$ are much larger 
than 1, the Maxwell distribution is adequate. If this condition is not satisfied by one or both, then
the relativistic distribution must be used. 
In Figure~\ref{Fig1} we show the
contours in the plane ($V_{\texttt{r}}$, $x_2$) of the 
relativistic distribution (\ref{P_v}), left-top panel,
and of the Maxwell distribution (\ref{Maxwell_vrel}), right-top panel.
We fix $x_1=500$ in the non-relativistic regime, and vary $x_2$ in the 
range (1, 40). The two distributions are very different in shape and absolute value.
Note in particular that $\mathcal{P}_{\texttt{r}}(V_{\texttt{r}})$ has a large peak in the 
relativistic region $V_{\texttt{r}} \gtrsim 0.8$, $x\lesssim 5$.
This peak is illustrated by plotting the relativistic $\mathcal{P}^{d}_{\texttt{r}}(V_{\texttt{r}})$, Eq.~(\ref{P_v_ii}), 
and Maxwell distribution with $\mu=m/2$ at $x=5$, first of 
the bottom panels of Figure~\ref{Fig1} where we study the case $m_1 =m_2 =m$. 
At small $x$ the distributions largely 
differ and become practically equal at $x\sim 100$.

\section{Mean value of the relative velocity}
An important quantity that characterizes the p.d.f. is the mean value of the relative velocity,
\begin{equation}
\langle V_{\texttt{r}} \rangle =\int_{0}^{1} d V_{\texttt{r}} \mathcal{P}_{\texttt{r}}( V_{\texttt{r}})
V_{\texttt{r}} .
\label{Vmean_def}
\end{equation}
It is convenient to use $\mathcal{P}_{\texttt{r}}(\gamma_{\texttt{r}})$, Eq.~(\ref{P_gamma}) and
$V_{\texttt{r}}=\sqrt{\gamma^2_{\texttt{r}}-1}/\gamma_{\texttt{r}}$. 
The integral in  Eq.~(\ref{Vmean_def}) then becomes
\begin{flalign}
\langle V_{\texttt{r}} \rangle =
\frac{X}{\sqrt{2} \prod_{i} K_2 (x_i)}    
\int_{1}^{\infty} d\gamma_{_{\texttt{r}}}
\frac{\gamma^2_{_{\texttt{r}}}-1}{\sqrt{\gamma_{_{\texttt{r}}}+\varrho}}
K_1 (\sqrt{2}X\sqrt{\gamma_{\texttt{r}} +\varrho}) .
\nonumber
\end{flalign}
We change variable to $y=(\gamma_{\texttt{r}}+\varrho)/(1+\varrho)$ and 
define
\begin{flalign}
c_2=(1+\varrho)^2, \;\;c_1=2\varrho(1+\varrho),\;\;c_0=\varrho^2-1.
\label{r's}
\end{flalign}
Using Eq.~(\ref{a}) and (\ref{r's}), after some algebra,  we find
\begin{flalign}
\langle V_{\texttt{r}} \rangle =\frac{\alpha}{2 \prod_{i} K_2 (x_i)}
(c_2 \mathcal{I}_2-c_1\mathcal{I}_1 +c_0 \mathcal{I}_0),
\end{flalign}
where the integrals
\begin{flalign}
&\mathcal{I}_2=\int_{1}^{\infty} dy \,
y^{\frac{3}{2}} K_1 (\alpha \sqrt{y}) =\frac{2}{\alpha} K_2 (\alpha)+\frac{4}{\alpha^2} K_3(\alpha),
\label{I2}
\\
&\mathcal{I}_1=\int_{1}^{\infty} dy \,
y^{\frac{1}{2}}  K_1 (\alpha \sqrt{y}) =\frac{2}{\alpha} K_2 (\alpha),
\label{I1}
\\
&\mathcal{I}_0=\int_{1}^{\infty} dy \,
y^{-\frac{1}{2}} K_1 (\alpha\sqrt{y}) =\frac{2}{\alpha} K_0 (\alpha),
\label{I0}
\end{flalign}
are calculated in the Appendix.

Using the recursion formulas of the Bessel functions we finally find
\begin{flalign}
\langle V_{\texttt{r}} \rangle=\frac{2}{\alpha}
\frac{c_2 K_3(\alpha) - c_{0} K_1 (\alpha)}{ K_2 (x_1) K_2(x_2)}.
\label{Vmean_x1_x2}
\end{flalign}
In the non-relativistic limit it gives the Maxwell's value 
\begin{equation}
\langle v_r \rangle=\sqrt{\frac{8 T}{\pi \mu}}=\sqrt{\frac{8}{\pi}}\sqrt{\frac{x_1 +x_2}{x_1 x_2}}.
\label{Vmean_Maxwell_x1_x2}
\end{equation}
It is easier to study the limit in the case $m_1 =m_2 =m$, the diagonal of the plane ($x_1$, $x_2$).
We have $\alpha=2x$, $c_2 =4$, $c_0=0$, thus formula (\ref{Vmean_x1_x2}) simplifies to
\begin{equation}
\langle V_{\texttt{r}} \rangle_d = 
\frac{4}{x}\frac{K_3(2x)}{K^2_2(x)}.
\label{v_rel_media}
\end{equation}
Using the  asymptotic expansions of $K_n (x)$ for $x\gg 1$
we find the non-relativistic expansion, 
\begin{equation}
\langle V_{\texttt{r}} \rangle_d \sim 4\sqrt{\frac{1}{\pi x}}\left[1-\frac{25}{16}\frac{1}{x}
+\frac{1305}{512}\frac{1}{x^2}+\mathcal{O}(x^{-3})\right].
\label{v_rel_expansion}
\end{equation}
To the lowest order coincides with the Maxwell value $\langle v_{{r}} \rangle_d =4\sqrt{{1}/{(\pi x)}}$  
as expected. 
In the ultra-relativistic limit, $x\ll 1$, we find
\begin{equation}
\langle V_{\texttt{r}} \rangle_d \sim 
1+\mathcal{O}(x^4),
\label{vrel_ultra}
\end{equation}
thus $\langle V_{\texttt{r}} \rangle$ tends to the velocity of light.

In the left panel of Fig.~(\ref{Fig2}) we show the contours of the mean  value (\ref{Vmean_x1_x2})
in the plane ($x_1$, $x_2$). 
At small $x$ it tends to 1 in both the "directions" $x_1$, $x_2$.
In the central panel we show  the contours of the Maxwell's value (\ref{Vmean_Maxwell_x1_x2}).
At $x_i <40$ the two mean values starts to differ significantly, and obviously the latter does not respect the
constraint $\langle v_r \rangle \leq 1$. 
In the right panel of Fig.~\ref{Fig2} we plot Eq.~(\ref{v_rel_media}). At $x\lesssim 0.7$,
$\langle V_{\texttt{r}} \rangle$
is already well approximated by the asymptotic value 1. In fact, as indicated by 
(\ref{vrel_ultra}), the first 
corrections to 1 is of the order $\mathcal{O}(x^4)$, thus very small. 
We find that the first three terms of the asymptotic expansion in Eq.~(\ref{v_rel_expansion})  
approximate well the exact function for $x\gtrsim 5$.
The  non-relativistic Maxwell value $\langle v_{{r}} \rangle =4\sqrt{{1}/{\pi x}}$
is always greater than the relativistic value $\langle V_{\texttt{r}}\rangle$ and it is a good approximation
for  $x >40$. 

\section{Summary and final remarks}

Guided by the principles of special relativity, Lorentz invariance and 
by the J\"{u}ttner distribution, we have found the probability distribution $\mathcal{P}_{\texttt{r}}( V_{\texttt{r}})$
of the relativistic relative velocity of binary collisions in a relativistic non-degenerate gas, Eq.~(\ref{P_v}), 
and an exact formula for the mean value of the relative velocity, Eq.~(\ref{Vmean_x1_x2}).
When at least one particle 
is relativistic, the Maxwell distribution is inadequate. Whenever a relativistic treatment is necessary,
the non-relativistic unit rate or thermal averaged cross section 
$\langle \sigma v_r \rangle=\int_{0}^{\infty} dv_r F_M (v_r) \sigma v_r$ can be replaced by the relativistic analogous
$\langle \sigma V_{\texttt{r}} \rangle=\int_{0}^{1} d V_{\texttt{r}} \mathcal{P}_{\texttt{r}}( V_{\texttt{r}})
\sigma V_{\texttt{r}}$. 

It is worth noting that a crucial step to derive the distribution is to not introduce, 
as usually done~\cite{Landau2,Cercignani,DeGroot},
the so called M\o{}ller velocity $\bar{v}={(1-\boldsymbol{v}_1 \cdot 
\boldsymbol{v}_2)}{{V}_{\texttt{r}}}=\frac{p_1 \cdot p_2}{E_1 E_2}{V}_{\texttt{r}}$,
but to maintain the factor $p_1 \cdot p_2/(E_1 E_2)$ that guarantees the invariance
of the product $n_1 n_2 (p_1 \cdot p_2)/(E_1 E_2)$ explicitly in the integral (\ref{Phi}).

One consequence of the present findings regards the thermal averaged cross section 
that appear in the calculation of the dark matter relic density.
After Ref.~\cite{GG}, it was accepted that in a relativistic framework, the velocity in 
$\langle \sigma {v} \rangle$ is the M\o{}ller velocity 
and for this reason often written in literature as "relative velocity".
The M\o{}ller velocity is not a fundamental quantity but it is derived from the relativistic relative 
velocity. 
We fully discuss this point in a separate work~\cite{paperSV}.

\acknowledgements

This work was supported in part by MultiDark under Grant No. CSD2009-00064 of the
Spanish MICINN  Consolider-Ingenio 2010 Program, by the
MICINN project FPA2011-23781 and by the Grant MICINN-INFN(PG21)AIC-D-2011-0724.

\appendix
\section{Integrals involving modified Bessel functions}

The integrals involved in the calculation of the mean relative velocity 
can be reduced to the known integrals~\cite{GR}
\begin{flalign}
&\int_{1}^{\infty} dz z^{\lambda} (z-1)^{\mu-1} K_\nu (a\sqrt{z})=\nonumber\\
&\;\;\;\;\Gamma(\mu) 2^{2\lambda-1} a^{-2\lambda} G_{1,3}^{3,0}
\begin{pmatrix}
\frac{a^2}{4} \Biggr\rvert
\begin{array}{c}
 0 \\
-\mu, \frac{\nu}{2}+\lambda,-\frac{\nu}{2}+\lambda,
\end{array}
\end{pmatrix},
\label{I_GR}
\\
&\int_{1}^{\infty}dz z^{-\frac{\nu}{2}} (z-1)^{\mu-1} K_\nu (a\sqrt{z})=\Gamma(\mu)2^{\mu}a^{-\mu} K_{\nu-\mu}(a),
\label{I_GR_simple}
\end{flalign}
where $G^{m,n}_{p,q}(x|\{a_1,...,a_n,...,a_p \},\{b_1,...,b_m,...,b_q\})$ is the generalized hypergeometric 
Meijer's $G$ function~\cite{GR}.
When one of the upper indexes is equal to one of the 
lower indexes the function is reduced to a simpler $G$ function, for example if $a_p=b_q=c$,
\begin{flalign}
G^{m,n}_{p,q}
\left(z\Bigg\rvert
\begin{matrix}{}
a_1,...,c\\
b_1,...,c 
\end{matrix}
\right)
=
G^{m-1,n}_{p-1,q-1}
\left(z\Bigg\rvert
\begin{matrix}{}
a_1,...,a_{p-1}\\
b_1,...,b_{q-1} 
\end{matrix}
\right).
\label{Prop_2_uguali}
\end{flalign}
The the modified Bessel functions are a particular $G$ function:  
\begin{flalign}
G_{0,2}^{2,0}\left(\frac{z^2}{4}\Bigg\rvert
\begin{array}{c}
 \\
\frac{\delta+\nu}{2},\frac{\delta -\nu}{2}
\end{array}
\right)=\frac{z^\delta}{2^{\delta-1}} K_\nu (z).
\label{K_2020}
\end{flalign}
Using the property (\ref{Prop_2_uguali}) we see that (\ref{I_GR_simple}) is
a particular case of Eq.~(\ref{I_GR}) with $\lambda=-\nu/2$.

The integral $\mathcal{I}_0$, Eq.~(\ref{I0}), follows directly form Eq.~(\ref{I_GR_simple}) with $\nu=\mu=1$.
The integral $\mathcal{I}_1$, Eq.~(\ref{I1}), follows from Eq.~(\ref{I_GR}) with 
$\lambda=1/2$, $\mu=1$, $\nu=1$ and reducing the resulting $G$ function
with the property (\ref{Prop_2_uguali}).
To calculate $\mathcal{I}_2$, Eq.~(\ref{I2}), we first note that Eq.~(\ref{I_GR}) 
with $\lambda=1/2$, $\mu=2$, $\nu=1$ gives
\begin{flalign}
\mathcal{J}&=\int_{1}^{\infty} dz 
z^{\frac{1}{2}} (z-1) K_1 (a\sqrt{z})
=\frac{1}{a} 
G_{1,3}^{3,0}
\begin{pmatrix}
\frac{a^2}{4} \Biggr\rvert
\begin{array}{c}
 0 \\
-2, 1, 0
\end{array}
\end{pmatrix}\crcr
&=\frac{1}{a}
G_{0,2}^{2,0}
\begin{pmatrix}
\frac{a^2}{4} \Biggr\rvert
\begin{array}{c}
  \\
-2, 1
\end{array}
\end{pmatrix}=\frac{4}{a^2}K_3 (a),
\end{flalign}
where we used (\ref{Prop_2_uguali}). Finally, $\mathcal{J}+\mathcal{I}_1=\mathcal{I}_2$ that gives
(\ref{I2}).


\begin{thebibliography}{100}
\bibitem{units}
Notations: we use natural units with $\hbar=c=k_B =1$ and Lorentz metric
$g=(1,-1,-1,-1)$. We use 'probability distribution' for
normalized probability density function (p.d.f.), $\int dx f(x) =1$.
\end{thebibliography}
\end{document}